\documentclass[a4paper,11pt]{article}
\usepackage[a4paper, margin=0.8in]{geometry}
\usepackage[utf8]{inputenc}
\usepackage{multicol}
\usepackage{wrapfig}
\usepackage{graphicx}
\usepackage[font=small,labelfont=bf]{caption}
\usepackage{comment}
\usepackage{epstopdf}
\usepackage{subfigure}
\usepackage{color}
\usepackage{siunitx}
\usepackage{fancyhdr}
\usepackage{lastpage}
\usepackage{amsmath,amssymb}
\usepackage{blindtext}
\usepackage{titlesec}
\usepackage[english]{babel}
\usepackage{xspace} 			
\title{\textbf{CERN AD/ELENA Antimatter Program}}
\date{\today}

\usepackage{authblk}

\author{
        R. Caravita\textsuperscript{1},        
        A. Cridland Mathad\textsuperscript{2},
        J. S. Hangst\textsuperscript{3},
        M. Hori\textsuperscript{4},
        B. M. Latacz\textsuperscript{2},
        A. Obertelli\textsuperscript{5},
        P.~Perez\textsuperscript{6},
        S. Ulmer\textsuperscript{7,8}\footnote{Corresponding author and ADUC chair, stefan.ulmer@cern.ch}, 
        E. Widmann\textsuperscript{9}
        \\
  on behalf of the Antiproton Decelerator User Community (ADUC)\footnote{The full author list is included in the Appendix}\\ \vspace{0.5cm}
      \textsuperscript{1}  TIFPA/INFN Trento, Italy\\
      \textsuperscript{2} CERN, Switzerland \\
      \textsuperscript{3} Aarhus Universitet, Denmark \\
      \textsuperscript{4} Imperial College London, United Kingdom \\
      \textsuperscript{5} Technische Universität Darmstadt, Germany \\
      \textsuperscript{6} IRFU, CEA, Universite Paris-Saclay, France \\
      \textsuperscript{7} Heinrich-Heine-Universität Düsseldorf, Germany \\
      \textsuperscript{8} RIKEN, Japan \\
      \textsuperscript{9} Stefan Meyer Institute, Austria }      

\begin{document}
\maketitle{}
\thispagestyle{empty}
\begin{abstract}
The CERN AD/ELENA Antimatter program studies the fundamental charge, parity, time (CPT) reversal invariance through high-precision studies of antiprotons, antihydrogen, and antiprotonic atoms. Utilizing the world-unique Antiproton Decelerator (AD) and the Extra Low Energy Antiproton (ELENA) decelerator, the program supports multiple groundbreaking experiments aimed at testing fundamental symmetries, probing gravity with antimatter, and investigating potential asymmetric antimatter/dark matter interactions. Some experiments focus on precision spectroscopy of antihydrogen, while others conduct the most precise tests of CPT invariance in the baryon sector by comparing proton and antiproton properties. Other efforts are dedicated to measure the ballistic properties of antihydrogen under gravity and performing antiproton-based studies of neutron skins in exotic nuclei. These efforts have led to major breakthroughs, including the first trapped antihydrogen, antihydrogen’s first gravitational acceleration measurement, and record-breaking precision CPT-tests in the baryon sector.

With continuous advancements in antimatter cooling, trapping, and transport, CERN’s program is opening new frontiers in fundamental physics. Future goals, described in this document and reaching to timelines beyond 2040, include further improving the precision of antimatter studies, developing transportable antimatter traps, and advancing our understanding of quantum field theory, gravity, and dark matter interactions.  Furthermore, new areas of hadron physics with antiprotons will be explored through studies of the Pontecorvo reaction, antineutron annihilation dynamics and hypernuclei decays. The CERN AD/ELENA Antimatter program remains at the forefront of experimental physics, pushing the limits of precision measurements to unravel the mysteries of the universe.
\end{abstract}
\newpage

\section{Scientific scope}
One of the fundamental pillars of the relativistic quantum field theories underpinning the Standard Model of particle physics is its invariance under simultaneous charge-conjugation, parity, and time-reversal (CPT symmetry). Remarkably, at the level of known physics, CPT invariance (CPT) stands as the only combination of discrete symmetry transformations observed as an exact symmetry of nature.
The CERN AD/ELENA Antimatter program is focused on testing CPT invariance by studying antiprotons, antihydrogen and antiprotonic atoms. Any result inconsistent with CPT symmetry would hint at new physics and potentially point to mechanisms that would allow to explain the matter-antimatter asymmetry observed on cosmological scales. 

CERN's AD and ELENA decelerators form a globally unique facility which produces low-energy antiprotons.  It serves four antimatter experiments simultaneously, with $100$ keV bunches of $1 \times 10^7$ antiprotons every two minutes, 24 hours a day from May to November. The AD user community consists of particle physics experiments employing atomic, molecular, and optical physics techniques which are ultimately making quantum limited measurements. The community currently consists of 6 active collaborations with about 350 scientists from 60 research institutes. 

CPT symmetry is tested in these experiments using antihydrogen spectroscopy across a wide frequency range (ALPHA, ASACUSA) and by studying the fundamental properties of the antiproton versus the proton (BASE).  Together with probing the Weak Equivalence Principle (WEP) with tests of the free-fall of antihydrogen (ALPHA, AEgIS, GBAR) and tests of the antimatter clock frequencies at different gravitational potentials (BASE).  This program is complemented by spectroscopy of antiprotonic helium to extract the antiproton to electron mass ratio (ASACUSA), scattering of antiprotons on unstable nuclei to study the neutron skin effect (PUMA, AEgIS) which could shed light on the nuclear processes within neutron stars and searches for coupling of dark matter with antimatter (BASE, ASACUSA).

The AD experiments have featured in the Physics World Top 10 Breakthroughs of the Year four times in the last 15 years.  Starting in 2010 with the first trapped antihydrogen (ALPHA, ASACUSA).  This was followed up with the first laser cooling of antihydrogen (ALPHA) and the sympathetic cooling of protons (BASE) in 2021, the first direct measurement of antihydrogen’s acceleration due to gravity (ALPHA) in 2023 and the first laser cooling of positronium (AEgIS) in 2024.  Other significant achievements include the most precise test of CPT invariance in the baryon 
sector with the proton/antiproton charge to mass ratio (BASE) and world leading limits on milli-charged dark matter (BASE).

\begin{wrapfigure}{r}{0.55\textwidth}
    \includegraphics[width=1.0\linewidth]{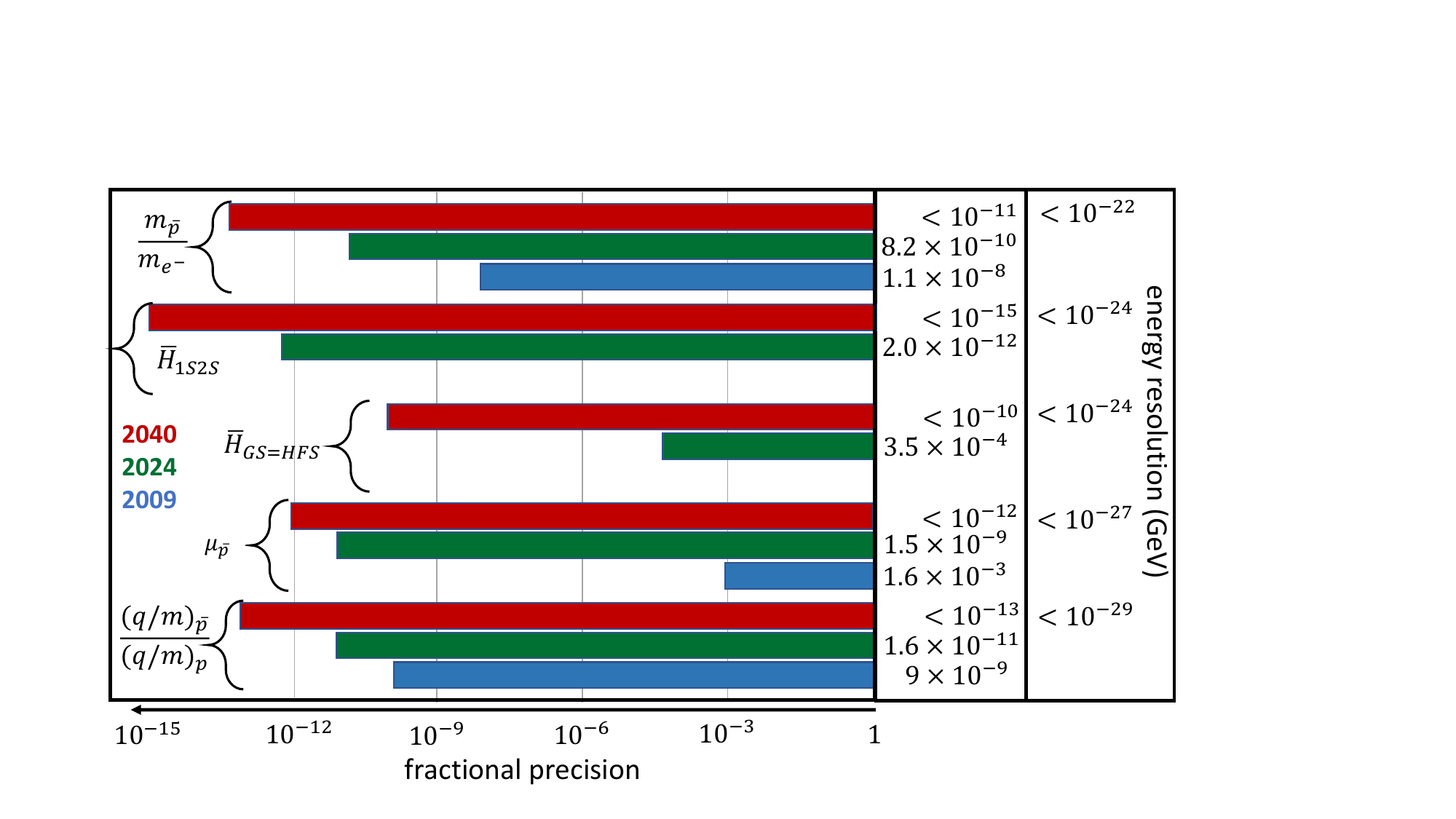}
    \caption{The current experiments' progress in precision achieved in the AD to date, along with the projected progress to be made by 2040.}
    \label{fig:AD_Progress}
\end{wrapfigure}


The AD experiments provide major breakthroughs every 5 - 10 years, shown by a 2 p.p.t. measurement of the 1S-2S transition in antihydrogen five years after the first trapping of anti-atoms (ALPHA), a six-orders of magnitude increase in the precision measurement of the antiproton magnetic moment in seven years (BASE) and a three orders of magnitude improvement in the antiproton to electron mass ratio in twenty years (ASACUSA).  More recent achievements include the production of an antihydrogen beam by charge exchange of antiprotons with a cloud of positronium (GBAR) and the first proton transport (BASE-STEP) in a transportable open access Penning trap. This document summarizes achievements, describes the current status, and outlines a future perspective beyond 2040 for the CERN AD/ELENA Antimatter program, some of which can be seen in figure \ref{fig:AD_Progress}.


\newpage

\section{AEgIS}
 
 The AEgIS experiment focusses on tests of gravity with antimatter, using a charge exchange process $\bar{p} + \text{Ps}^* \rightarrow \bar{\text{H}} + e^-$ between antiprotons and Rydberg atoms to form a pulsed beam of antihydrogen atoms. Thanks to the expertise gained in implementing laser-controlled pulsed formation of $\bar{\text{H}}$~\cite{AEgis_Hbar_2019} and the related controlled formation and manipulation of excited states of positronium~\cite{AEgIS_laser_excitation_Ps}, a number of further promising physics topics have been incorporated into AEgIS' physics program. The developments required to cover this initial program and the longer term research topics  are separated into near-term and longer-term prospects, all of which are feasible within the AEgIS experimental setup and are actively being pursued.

 \subsection{Near and medium term physics topics (beyond LS3 and up to LS4)}

 \begin{itemize}
     \item Gravity with antihydrogen: a first measurement relying on pulsed antihydrogen formation of a horizontally travelling beam using Moir\'e interferometry is feasible by LS3, a more sensitive measurement will be possible in the post-LS3 period
     \item Laser cooled positronium: utilising laser cooling of positronium~\cite{AEgIS_laser_cooling_Ps}, a test of the WEP and first precision Ps spectroscopy can be envisaged on a similar time scale
     \item Antiprotonic atoms: pulsed formation of antiprotonic (Rydberg) atoms with a view to carrying out laser spectroscopic studies of $\bar{p}$--atoms other than $\bar{p}$He are under way~\cite{AEgIS_antiprotonic_atoms, PhysRevC.109.064320}
     \item Nuclear physics: annihilation of antiprotons on trapped atoms~\cite{AEgIS_nuclear_fragments} and the study of the resulting trapped nuclear fragments forinter alia mass measurements of radioisotopes that are not produced at ISOLDE
 \end{itemize}

\subsection{Medium to long term physics topics (up to and beyond LS4)}

\begin{itemize}
    \item Precision testing of QED: spectroscopy of highly-charged hydrogen-like ions which contain an antiproton.  These are formed via controlled formation of antiprotonic atoms, followed by the trapping of the resulting fully stripped nuclear fragments and dressing these with an electron or an antiproton through charge exchange processes~\cite{AEgIS_HCI}. Precision testing of QED and of atomic constants can also be achieved with spectroscopy of laser-cooled Ps
    \item Antiprotonic molecules: incorporation of antiprotons in Rydberg states in molecules and molecular ions opens the door to searching for an antiproton-EDM
    \item Anti-deuteronic atoms: should sufficient numbers of low energy $\bar{d}$ become available the formation and precision spectroscopy of anti-deuteronic atoms tests atomic physics with a spin-1 'heavy electron'~\cite{AEgIS_deuteronic_atoms}
    \item Antineutrons: controlled production of very low-energy antineutrons from antiproton annihilations in a nuclear charge exchange process allows the antineutron-proton and antineutron-neutron annihilation processes to be probed at threshold
    \item Search for dark matter: $\bar{p}$~-$^3$He annihilations at rest enable a unique search for the uuddss sexaquark dark matter candidate~\cite{AEgIS_sexaquark}
\end{itemize}

\subsection{Summary and outlook for the coming decade}
AEgIS has established the techniques (charge exchange, laser control, precision detectors, ion injection) needed to tackle its initial goals (gravity measurements with antihydrogen and spectroscopy of positronium). It is now laying the foundations for a wide range of further physics topics building on these pulsed atom control techniques such as gravity tests with multiple pure antimatter or mixed matter-antimatter systems, precision tests of QED in novel systems, nuclear physics, searches for dark matter and studies of molecular systems that incorporate antimatter. Achieving these reachable goals relies on the long term availability and support for CERN's unique antiproton facility beyond LS4.

\newpage

\section{ALPHA}

The ALPHA experiment has been actively studying the properties of the antihydrogen
atom since 2010, when the first anti-atoms were trapped \cite{andresen_trapped_2010} in our magnetic multipole trap. Since
then, we have observed many optical \cite{ahmadi_observation_2017-1, ahmadi_observation_2018, ahmadi_investigation_2020} and microwave transitions  \cite{amole_resonant_2012, ahmadi_observation_2017} in antihydrogen, and we have performed the only gravitational \cite{anderson_observation_2023} experiment to date with an antimatter atom. Our physics
program over the next two decades will continue to focus on tests, with ever-improving
precision, of CPT invariance and the gravitational Weak Equivalence Principle.

\subsection{Spectroscopy of antihydrogen}

We have long prioritized the measurement of the 1S-2S transition in antihydrogen. The
measured precision of this frequency in hydrogen \cite{parthey_improved_2011} is of order $10^{-15}$; our current published result \cite{ahmadi_characterization_2018}
is of order $10^{-12}$, with greatly improved results soon to follow. Recently, we have demonstrated
laser \cite{baker_laser_2021} and adiabatic cooling \cite{the_alpha_collaboration_adiabatic_2024} of trapped antihydrogen, and we are able to accumulate 20,000
atoms overnight, using a new technique involving positrons cooled by interaction with laser-cooled Be$^+$ ions \cite{baker_sympathetic_2021}. We thus expect a steady improvement in this, the most precise, direct
comparison between matter and antimatter atom. The next major innovation here will be to introduce
hydrogen into our magnetic trap to allow direct, in situ comparison of the 1S-2S and other transitions.
This will be our priority during the upcoming Long Shutdown 3 (LS3) at CERN and promises
a decade of unique, fundamental physics results. We have a primary time standard, a Caesium
fountain clock, in our lab for direct frequency metrology. We see no fundamental obstacles to
obtaining the same precision demonstrated in hydrogen. Our new accumulation technique \cite{ahmadi_antihydrogen_2017}
allows for rapid study of systematics and offers new possibilities for sidereal measurements.
Other transitions have already been studied or will soon be examined. The Lyman Alpha
(1S-2P) transition is the basis for laser cooling and has been extensively studied. The ground
state hyperfine splitting has already been observed and quantified using microwaves and its
precision is currently being improved. We will soon (LS3) implement changes to allow direct
excitation of the NMR transition in trapped antihydrogen. We have recently observed the 2S-2P and the 2S-4P lines in the ALPHA-2 machine. The necessary laser hardware for the 2S-3S
transition is on-site and being commissioned. ALPHA thus represents a comprehensive suite of
experiments to characterize antihydrogen and to extract fundamental quantities such as the
antiproton charge radius and the anti-Rydberg constant in the coming years. We also are currently
upgrading the spectroscopy apparatus (the ALPHA-3 project) to include fluorescence light
detection and improved optics for the 1S-2S study.

\subsection{Antimatter and gravitation}

ALPHA demonstrated in 2023 that antihydrogen is gravitationally attracted to the
Earth, and we measured the acceleration to a precision of about 20\%. The ALPHA-g machine
is poised to continue and to improve such measurements, which were already reproduced in
our run of 2024. We expect a precision of 1\% or better is possible in the existing machine, and
we are currently planning upgrades (LS3 and beyond) that will include extraction of the
antihydrogen from the magnetic trap and atomic interferometry to reach much higher levels of
precision. With our new techniques to laser cool antihydrogen and to accumulate tens of
thousands of atoms using laser cooled Be$^+$ ions to cool the positrons, we are confident that the
future of gravitational studies with antimatter is extremely promising.

\subsection{Summary}

Physics with antihydrogen is now routine in ALPHA. Our trapped antihydrogen atoms
can currently be cooled to a mean energy of order 1 mK (in temperature units). With such a
source of more than $10^4$ simultaneously trapped atoms, the horizon for antihydrogen studies is
essentially limitless. It is fair to say that the core physics goals of the original Antiproton
Decelerator physics program are now in hand and will be easily surpassed in the coming
decades.
\newpage

\section{ASACUSA}
The Atomic Spectroscopy And Collisions Using Slow Antiprotons (ASACUSA) collaboration studies the fundamental symmetries between matter and antimatter using precision spectroscopy of atoms containing an antiproton. These can be hybrid matter-antimatter atoms (antiprotonic helium) or pure antiatoms (antihydrogen). The program also includes antiproton annihilation and collision studies. \vspace{0.025cm} \\
\textbf{{Antihydrogen ground-state hyperfine structure:}} 
The ASACUSA antihydrogen program aims to perform precision spectroscopic studies to investigate the properties of the antimatter atoms and place constraints on beyond standard model physics. To do this, we produce an antiatomic beam which can be extracted from the strong magnetic fields necessary to trap charged particles in the antihydrogen formation region \cite{kuroda_source_2014, Kolbinger:2021}. Initially, we plan to measure the ground state hyperfine structure (GS-HFS) using a Rabi spectroscopy method, which will achieve p.p.m. precision. \\ 
The AD-ELENA upgrade, along with new techniques for plasma cooling and purification, enabled us to achieve a factor of 100 increase in beam intensity. Beam parameters such as intensity, spatial distribution, ground state fraction, polarization, and velocity will be further improved with work on plasma properties to enable higher precision spectroscopic measurements. In parallel, work on focusing and decelerating the formed beam as well as alternative beam formation mechanisms will take place. Upgrading the Rabi spectroscopy setup to a Ramsey type will improve the precision of the GS-HFS measurement from p.p.m. to 50 p.p.b. Spectroscopic measurements of other states of antihydrogen may be performed in the beam using microwave, or in the future once a fine focus has been achieved, laser interactions. Depending on the measurement, these may be sensitive to properties such as the antiproton radius or an anti-Rydberg constant. We hope to reach the same precision that we achieve using a hydrogen beam and our spectroscopy apparatus \cite{nowak_cpt_2024} with antihydrogen ($<$p.p.b.). 
The ultimate precision in laser and hyperfine spectroscopy will require the usage of a fountain as described in \cite{comparat2024experimental}, and is foreseeable after LS4. \vspace{0.025cm} \\
\textbf{{Antiprotonic helium and antiproton-to-electron mass ratio:}}
Antiprotonic helium is a three-body atom composed of a helium atom with one of its electrons replaced by an antiproton occupying a Rydberg orbital. It is unique among the ``half-matter, half-antimatter" hadronic exotic atoms and complements antilepton-antihadron antihydrogen in tests of quantum electrodynamics (QED) at the highest precision, owing to the fact that it survives for anomalously long times of $\ge 10$ $\mu{\rm s}$. ASACUSA has employed quantum optics techniques to measure the transition frequencies of the atom and compared the results with QED calculations, thereby determining the antiproton-to-electron mass ratio as $M_{\overline{p}}/m_e=1836.1526734(15)$ \cite{hori_buffer_2016,hori2011two}. This data has also been used to set upper limits on any hypothetical fifth forces that may arise between hadrons and antihadrons acting at length scales of $10^{-11}$ m and on exotic velocity and spin-dependent, semi-leptonic forces that may arise between antihadrons and electrons due to axions. \\
ASACUSA currently attempts to improve the precision of $M_{\overline{p}}/m_e$ and this CPT consistency test by a factor of 100 by employing the technique of two photon laser spectroscopy. This technique allowed the first laser spectroscopy of an atom containing a meson at the Paul Scherrer Institute \cite{hori_laser_2020} where efforts are underway to improve the precision of the charged pion mass $M_{\pi}$. Laser spectroscopy of a kaonic atom to increase the precision of the charged kaon mass $M_K$ by $\ge 1000\times$ is also under discussion. Surprising behaviours of antiprotonic helium suspended in superfluid helium \cite{soter2022high} may allow antiprotons to probe condensed matter effects. \vspace{0.025cm} \\
\textbf{{Antiproton annihilation and collision experiments:}}
Using a continuous antiproton beam from our catching trap, we measured ionization cross sections for helium, argon, and molecular hydrogen at $10\mbox{--}20\,\mathrm{keV}$. The antiproton serves as an ideal (slow, repulsive, no exchanged electrons) charged probe of the electron distributions and dynamics in the target atoms and molecules. We aim to extend these measurements to $1\,\mathrm{keV}$ and below. The antiproton beam is also directed to thin foils to benchmark models of an antiproton annihilation at rest \cite{amsler_antiproton_2024}. To refine antiproton-nucleus interaction studies, we plan to measure annihilation cross-sections at 100 keV with ELENA in mini-bunch mode \cite{aghai_limits_2021}. \vspace{0.025cm} \\
\textbf{{Future proposals:}} Future AD upgrades may enable studies at 5.3 MeV and above, covering annihilation and elastic scattering \cite{aghai_measurement_2018}. For more details, see Section~\ref{New_Proposals}, where we also propose to measure the Pontecorvo reaction \cite{Venturelli_pontecorvo_2025} and perform antimatter interferometry.
\newpage

\section{BASE}

The BASE collaboration is operating advanced Penning trap systems to compare the fundamental properties of protons and antiprotons with ultra-high precision.  We have compared proton and antiproton charge-to-mass ratios with a fractional accuracy of 16 parts-per-trillion \cite{borchert202216}, which constitutes the most precise test of CPT invariance in the baryon sector to-date. The study also enables us to interpret our measurements as clock comparisons, and to set limits on the weak equivalence principle. Another part of the BASE physics program is high-precision measurements of the proton and the antiproton magnetic moments. While we have determined the proton magnetic moment with an accuracy of 300$\,$p.p.t$.$ \cite{schneider2017double}, our 1.5$\,$p.p.b$.$ measurement of the antiproton magnetic moment \cite{smorra2017parts} improved the previous best result by more than a factor of 3000. This measurement also enabled us to set direct limits on antimatter dark matter coupling \cite{smorra2019direct}, which improved previous astrophysical limits by more than five orders of magnitude.\\
\textbf{{Status and Mid-Term Plans:}}
With a considerable revision of our apparatus we have demonstrated very recently, that coherent proton/antiproton magnetic moment measurements are possible.  In these studies we have sampled g-factor resonances with line-widths 16 times narrower than in \cite{smorra2017parts} and at 1.5 times higher signal inversion. This will enable antiproton magnetic moment measurements with at least 10-fold improved accuracy, and has the potential to study antiproton/axion couplings with drastically enhanced sensitivity. After the current magnetic moment measurements, new charge-to-mass ratio measurements are planned, that will be based on simultaneous measurements on two particles in one trap. Here, we expect do determine the $\text{p}/\bar{\text{p}}$ charge-to-mass ratio with  p.p.t$.~$accuracy. \\
\textbf{{Offline Antiproton Experiments:}}
During accelerator up-time, magnetic noise from AD/ELENA limits our measurements. To overcome this, we have developed an antiproton reservoir trap \cite{smorra2015reservoir}, allowing measurements during the facility’s yearly technical stop. However, the three-month shutdown is insufficient to create significant progress in high precision studies. To address this, we plan to relocate experiments to a dedicated low-noise precision laboratory. Recently, we have achieved a key milestone by demonstrating the first transport of protons out of CERN’s antimatter factory \cite{smorra2023base}.\\
\textbf{{Long Term Experiment Program:}}
Currently, we are developing antiproton receiver experiments in offline laboratories at HHU-D\"{u}sseldorf, and are planning to relocate the BASE-CERN experiment to an offline laboratory at CERN. This strategy will allow synchronized charge-to-mass ratio and magnetic moment measurements in instruments with horizontal and vertical geometry. In addition, this will also allow independent measurements of charge-to-mass ratios and magnetic moments in apparatuses with different features, for enhanced measurements on both numbers. Having established $\bar{\text{p}}$-transport and spectroscopy in offline laboratories, at least 100-fold improved measurements of the $\bar{\text{p}}$-fundamental constants are in reach. The construction of new experiments and establishing $\bar{\text{p}}$-transport, iterations and refinement of the measurement techniques will range until 2040, at least. \\
Having established these technologies, the entire BASE infrastructure -- consisting of experiments at Mainz, Hannover, Heidelberg and D\"{u}sseldorf -- will profit from these developments. In the long term, we are planning to install a network of antiproton experiments driving our future measurements to considerably higher accuracy, and establishing new technologies. For example, the BASE Mainz experiment has demonstrated sympathetic cooling of single protons to record temperatures \cite{bohman2021sympathetic}, and in the BASE-Hannover experiment
p-based quantum-logic spectroscopy \cite{cornejo_quantum_2021} is being developed that has potential to be applied to the $\bar{\text{p}}$. These quantum-resolved methods promise to drastically enhance the sampling rate of our experiments.
In addition, we are developing experiments to investigate the deuteron magnetic moment, to have methods available if CERN decides to establish antideuteron production, and we are planning to implement a dedicated antiproton lifetime experiment to improve the current best direct limits by a factor of $\approx$1000 \cite{sellner2017improved}. \\
\textbf{{Exotic Projects}:}
The BASE spin state detection trap exhibits heating rates orders of magnitude lower than any previously characterized trap \cite{borchert2019measurement}. Its high sensitivity enables the study of charge and mass ranges in which milli-charged dark matter can exist, setting world-leading limits in certain regimes \cite{budker2022millicharged}, with potential for improvement. Additionally, the ultra-sensitive detectors used for single trapped antiprotons can serve as axion and axion-like particle detectors \cite{devlin2021constraints}. Setting up a dedicated experiment aiming to enhance previous limits by at least a factor of 100 is under consideration.

\newpage

\section{GBAR}

The goal of the GBAR collaboration is first to produce the anti-hydrogen ion (one antiproton and two positrons) as an intermediate state that can be cooled sympathetically by laser cooled matter ions such as $\mathrm{Be^+}$. Once cooled the anti-ion can be photo neutralised near threshold in order to
induce a minimal kick to the then ultra cold anti-atom~\cite{GBARProposal}, while keeping control of the height of the free-fall and the
distribution of initial velocities. The anti-atom free fall can then be observed in a
classical way over heights of the order of 10~-~20 cm as a first step and then over heights smaller
than a mm where quantum effects from surface forces come into play. In particular, the anti-atoms
can be reflected by the Casimir-Polder potential, leading to the formation of gravitational quantum
states above a surface. This will allow us to reach a relative precision on the gravitational acceleration of antihydrogen of order
$10^{-5}$~\cite{quantumreflection}. The anti-ion will be a novel tool for antimatter experiments and has yet to be produced. This will allow for new exciting future perspectives in this field such as optical trapping of ultracold antihydrogen atoms~\cite{Crivelli:2017fgg} and production of molecular ions~\cite{myers2018cpt} which will result in an unprecedented sensitivity to test Lorentz and CPT symmetry.

\subsection{Status}
In the GBAR scheme the anti-ion is produced in a series of two consecutive charge exchange reactions on positronium. The first reaction with antiprotons $\mathrm{\overline{p} + Ps \rightarrow \overline{H} + e^-}$ is followed by the antihydrogen atoms reacting with positronium  $\mathrm{\overline{H} + Ps \rightarrow \overline{H}^+ + e^-}$. This requires us to produce and focus large quantities of antiprotons and positrons, typically of the order of $10^7$ and $10^{10}$ respectively, for each interaction to produce about one anti-ion. The experiment is now able to trap $\mathrm{5 \times 10^6 \; \overline{p}}$ from ELENA and $\mathrm{6 \times 10^8 \;e^+}$ every 2 minutes. The positrons are converted into ortho-positronium after implantation into a nanoporous silica film. Record numbers of $\mathrm{7 \times 10^7 \; \overline{p}}$ trapped antiprotons in 35 minutes and close to $\mathrm{10^{10} \;e^+}$ trapped positrons in 30 minutes have been achieved. 
The production of antihydrogen through the first reaction was demonstrated in 2022~\cite{GBARHbarprod}.
Following the upgrades implemented during 2023, the production rates increased by a factor of 30 compared to 2022, with the potential for a further 2 orders of magnitude improvement once transport to the interaction point will be optimised. 

\subsection{10-year plan}
Latest developments in few-body atomic collision theory provide cross sections for antihydrogen atom and ion formation from
positronium, but with large discrepancies between models. An experimental measurement of these cross sections is planned for 2025 using the GBAR set-up and a pulsed hydrogen beam~\cite{GBARSPHINX}. 
If the cross section is found to be large enough, GBAR will be able to perform its physics program on the WEP
in the years after the CERN Long Shutdown 3 (LS3, until 2028). The first years will be used to demonstrate and master the steps of $\mathrm{ \overline{H}^+}$ sympathetic cooling and photodetachment at threshold, before the classical free fall measurement itself reaches a 1 \% precision. This will be followed by the first experimental evidence of quantum reflection of cold antihydrogen atoms from the surface and the first observation of gravitational quantum states (GQS). On a reasonable timescale during the CERN Long Shutdown 4 (2034), GBAR has the potential to shift toward GQS spectroscopy and interferometry, that will enable measurements at $10^{-5}$ precision at least.

In addition to the main physics program on gravitation, GBAR will also complement measurements started before LS3, namely the Lamb shift for antihydrogen atoms in a magnetic field free environment~\cite{GBARLambshift},
cross-sections for the production of antihydrogen atoms and ions via charge exchange with excited positronium.

The in-flight spectroscopy program, which can run parasitically at the same time as $\mathrm{ \overline{H}^+}$ production, may be extended to other microwave (MW) transitions in antihydrogen such as the fine structure which would provide a complementary test of Lorentz and CPT invariance. 
The in-flight MW spectroscopy can also be extended to excited states spectroscopy for a complemental test of the CPT symmetry.
Looking further ahead, if antideuterons become available at the AD, the GBAR set-up can be upgraded with a scheme to recycle the antideuterons and the same physics program could be pursued on antideuterium~\cite{GBARantiD}.

\newpage
\section{PUMA}

Neutron excess in nuclei results in the development of a neutron skin at the nuclear surface. Neutron skins demonstrate the properties of effective nuclear forces emerging from QCD and are related to the nuclear shell structure, correlations at the nuclear surface and the nuclear equation of state, which drives the physics of neutron stars. Despite tremendous efforts in determining neutron skin thicknesses in stable closed-shell nuclei, our experimental knowledge remains limited due to the difficulty in investigating the neutron density distribution. 

Low-energy antiprotons can be captured onto an orbital in the Coulomb field of a nucleus and form an antiprotonic atom. They offer a very unique sensitivity to the neutron and proton densities at the annihilation site, i.e. in the tail of the nuclear density where the neutron skin develops. Such studies are the core of the antiProton Unstable Matter Annihilation (PUMA) experiment \cite{aumann2022}. 

PUMA focuses on the detection of charged particles, mostly pions, following the annihilation. The total charge of the produced pions, as well as the multiplicity of charged pions detected, will be used to extract the number of annihilations with neutrons and with protons. The main uncertainty of the measured ratio comes from the correction of the final state interactions of the produced pions with the residual nucleus. These studies with radioactive ions were first proposed by Wada and Yamazaki in 2004 \cite{wada2004} and it has been demonstrated that a precision of 10\% can be achieved for $10^4$ annihilations. The concept has been further developed and implemented for the first time in the PUMA experiment.

PUMA is composed of Penning traps in a 4-T solenoid. The trap in which ion-antiprotons annihilations will be  performed is surrounded by a time-projection chamber for charged-pion detection. Experiments with stable isotopes will be carried out at ELENA, while experiments with short-lived nuclei will be performed at ISOLDE. For this reason, the experiment is made transportable; antiprotons will be trapped, stored and then, the experiment will be moved to ISOLDE for measurements.

\subsection{Status and 10-year Plan}

The experiment was accepted in 2021. The experimental setup is close to completion and it will be fully installed at ELENA in 2025. In the first year, we aim at demonstrating the principle of the experiment with the storage and transport of antiprotons ($10^7$ antiprotons as a first milestone) to ISOLDE and the efficient mixing of antiprotons and isotopes.
\\

\noindent At ELENA, we will:
{\renewcommand\labelitemi{}
\begin{itemize}
\item (i) determine the neutron-to-proton annihilation ratio for light systems: p, d, $^3$He,$^4$He and stable nuclei along the oxygen, neon and xenon isotopic chains to investigate the mass and isospin dependence of the neutron-to-proton annihilation ratio, 
\item (ii) characterize the neutron skins of the stable isotopes $^{40-48}$Ca, $^{112-124}$Sn and $^{206-208}$Pb by use of a to-be-developed laser-ablation ion source.
\end{itemize}
}

\vspace{0.3cm}

\noindent At ISOLDE, we will explore the dependence of the observables with the neutron number, focusing on the following cases: 
{\renewcommand\labelitemi{}
\begin{itemize}
\item (i) $^{6,8}$He, $^{11}$Li known as neutron halos, 
\item (ii) $^8$B, $^{17,18}$Ne candidates for proton halos or proton-skin nuclei, 
\item (iii) $^{26-30}$Ne, $^{28-33}$Mg to investigate the role of deformation and search for p-wave halos, 
\item (iv) $^{14-22}$O, exotic Xe and Sn isotopes to investigate the evolution of neutron skins with the neutron number.
\end{itemize}
}
\vspace{0.5cm}

The most difficult cases (short lived and low radioactive ion intensity, such as $^{11}$Li with  a lifetime of 9 ms) and a beam-on-target intensity $\sim 10^3$ particles per second, will require PUMA to achieve its nominal performance. 


\section{New Proposals}
\label{New_Proposals}

In addition to the long range plans highlighted within the existing experiments, new physics ideas such as antiprotonic atom X-ray spectroscopy (PAX), spectroscopy of hypernuclei (HY$\mathrm{\overline{P}}$ER), antihydrogen molecular ion (collaboration of many groups into which members from ALPHA, GBAR, BASE, HHU, MPIK and many others will be involved), study of properties of antideuterons (AEgIS, ASACUSA, BASE, EXEQT and GBAR), collision experiments using continuous beams (ASACUSA), and low energy antineutron physics, are discussed in this section. Some of these ideas are already in the preparation phase, including successful or ongoing grant applications.

\subsection{Antiprotonic Atom X-ray Spectroscopy}

The PAX project (antiProtonic Atom X-ray spectroscopy) seeks to perform, for the first time, the highest-precision x-ray spectroscopy in anti-protonic atoms. It depends on a new paradigm of focusing on transitions between circular Rydberg states in exotic atoms to probe higher-order strong-field QED free from nuclear uncertainties \cite{paul2021testing}. The realization of PAX is made possible based on the availability of low-energy antiproton beams at ELENA, notably the recently demonstrated Multiple Mini Bunch Extraction (MMBE), and newly available Transition Edge Sensor (TES) microcalorimeter detectors, capable of achieving $1\times 10^{-4}$ intrinsic resolution while maintaining a quantum efficiency of 0.4-0.8~\cite{baptista2025towards}.

The PAX physics program rests on a new paradigm of focusing on transitions between circular Rydberg states in exotic atoms \cite{paul2021testing}. The heavy mass of the antiproton leads to a small Bohr radius in the antiprotonic atom, leading to orders-of-magnitude stronger electric fields that then enhance the QED contributions in the transition energies. A subset of transitions between circular Rydberg states can be found where the second-order QED effects, notably vacuum polarization effects, become measurable with new techniques, while nuclear uncertainties are negligible. PAX will measure these transitions at the $1\times 10^{-5}$ accuracy level with a novel large-area TES detector developed by the quantum sensing division of NIST (USA), and probe second-order QED effects never before studied in a variety of antiprotonic atoms made with gaseous targets. 

To achieve the ultimate precision with these quantum sensing x-ray detectors, the instantaneous photon rate on the detector must be made as low as possible. Thus, the MMBE antiproton extraction at ELENA is essential, and it is estimated that the project would be able to attain an order of magnitude higher accuracy if a true slow-extraction antiproton beam were available. When highest accuracy measurements are achieved, the PAX methodology may allow to place additional constraints on new baryonic interactions with decays to the dark sector \cite{liu2025probing}.


\subsection{Antihydrogen Molecular Ion}
Ro-vibrational spectroscopy of the antihydrogen molecular ion, $\bar{\text{H}}_2^-$ \cite{myers2018cpt}, and its comparison to the hydrogen molecular ion \cite{schenkel2024laser} offer a promising avenue for testing CPT invariance with fractional resolutions on the order of $10^{-17}$ \cite{schiller2024prospects}. These measurements are highly sensitive to the antiproton/proton and electron/positron mass ratios, (anti)proton–(anti)proton interactions \cite{alighanbari2023test}, exotic physics \cite{germann2021three}, and multiple coefficients of the Standard Model Extension \cite{shore2024lorentz,vargas2025prospects}. Recently, laser spectroscopy of $\text{H}_2^+$ has reached an experimental uncertainty of less than $1\times10^{-11}$ \cite{Alighanbari} serving as a strong foundation. However, synthesizing this system remains a significant challenge, and even if successful, only a very limited number of particles are expected to be available. The BASE collaboration has demonstrated measurements with parts-per-trillion resolution while operating at a particle consumption rate of just one particle per two months, which are potentially promising techniques to be applied to species at very low production rate.\\ 
Extending the antihydrogen synthesis capabilities of the ALPHA collaboration \cite{ahmadi_antihydrogen_2017}, while in parallel investigating possibilities to synthesise the antimolecular ion within GBAR -- either by production of hot antimolecular ions in a merged beam configuration from two antihydrogen atoms or from one $\bar{\text{H}}^{+}$ with one antiproton, or the formation of cold antimolecular ions in a trap from cold antiprotons and sympathetically cooled $\bar{\text{H}}^{+}$ ions -- production paths might become available in the distant future. Combining these studies with the specialized knowledge of hydrogen molecular ion spectroscopy, such as e.g. available at Heinrich Heine University Düsseldorf \cite{schiller2024prospects}, and advanced methods for studying these ions in Penning traps (BASE), the resulting consortium would have a potential prospect of investigating $\bar{\text{H}}_2^-$ in the future. In the very far distant future, this could even lead to  possible spectroscopy of the $\overline{\text{HD}}^{-}$ molecule.

\subsection{Hypernuclei Physics}

Observable hadronic matter is mostly composed of $u$ and $d$ quarks. The strangeness degree of freedom is expected to play an important role in dense nuclear matter in the form of hyperons, baryons with at least one valence strange quark. Hyperons can form bound nuclear systems with nucleons and create short-lived hypernuclei which decay on the time-scale of the weak interaction (typically $\sim$200~ps). Our knowledge of nuclear matter is so far restricted to precision nucleon scattering and nuclear data while a generalisation to other quark flavours lacks precision data.

THe HY$\mathrm{\overline{P}}$ER project will be based on the production of hypernuclei following the capture and annihilation of low-energy antiprotons. HY$\mathrm{\overline{P}}$ER will explore the terra incognita of the strange nuclear landscape. The measured precision ground-state properties and spectroscopy of single-$\Lambda$ hypernuclei along isotopic chains, from neutron deficient to neutron rich, will give access to the isospin properties of the many-body interactions involving $\Lambda$ hyperons,  and thus to the role of strangeness in the nuclear equation of state and in neutron stars.

HY$\mathrm{\overline{P}}$ER will measure the binding energy of 200 hypernuclei with a resolution better than 300~keV, reaching a precision of few 10 keV for most of them, well below most of available data often suffering from systematic errors, in particular when obtained from emulsion pictures. HY$\mathrm{\overline{P}}$ER will allow to investigate systematically the isospin degree of freedom by producing new hypernuclei with mass $A>10$, where the density inside the nucleus reaches the nuclear saturation mass density $\rho \sim \rho_0$. This is also new and such new structure information will become a reference to constrain the hyperon-nucleon (YN) and three-baryon (YNN) interactions. As an example, the systematics of binding energies along the oxygen isotopic chain will be extended from $^{13}_{\Lambda}$O to $^{21}_{\Lambda}$O. This new structure information should become a reference to constrain and develop chiral effective field theories (EFT) interactions involving $\Lambda$ hyperons and nucleons. A fundamental difference between $\Lambda$s and nucleons is that the first excited state of the $\Lambda$, the $\Sigma$ hyperon, is at an excitation energy of only 77 MeV, while the first excited state of the nucleon, the $\Delta$ resonance, is at about 300 MeV: neighboring nucleons at saturation density will have a much stronger effect on the polarisation of the $\Lambda$, in the form of $\Lambda-\Sigma$ coupling, than for the nucleon. This feature in hypernuclei is a magnifying glass to in-medium baryon excitations, i.e., effects in nuclei which challenge chiral EFT at a certain level of precision for a given resolution scale. \\
\subsection{Antineutron/Antideuteron Physics}

One strategy to investigate the fundamental properties of the antineutron is to study antideuteron systems, which are easier to produce and handle experimentally. Its production at AD/ELENA is currently being studied and has been estimated at around 400 antideuterons per 2-minute cycle, opening up a wealth of new experimental opportunities in fundamental physics. In this section we summarize the main ideas proposed by the community.



The first concept is to perform mass spectrometry and magnetic moment measurements of the antideuteron. Using the techniques applied in BASE for proton/antiproton charge-to-mass ratio comparisons \cite{borchert202216}, $\text{d}/\bar{\text{d}}$ comparisons at the parts-per-trillion level would be immediately feasible. 
Measurements of the $\bar{\text{d}}$ magnetic moment constitute a dramatic challenge, since the continuous Stern-Gerlach Effect that is used for spin detection scales as $\mu/m$, where $\mu$ and $m$ are the magnetic moment and the mass of the particle. A deuteron magnetic moment measurement would require the development of an apparatus that is, compared to antiproton experiments, six times more sensitive with respect to magnetic moments, while the antiproton magnetic moment techniques are already at the principal technical limit. The possibility of the development of such an apparatus is currently under investigation by the BASE collaboration.

Another idea would be to perform a spectroscopy program for anti-deuterium atoms within AEgIS Penning traps or in-flight in a magnetic field free environment with the upgraded GBAR apparatus, which offers an unique opportunity of recycling antideuterons. According to the simulations, in a deuterium equivalent production mechanism of antihydrogen in GBAR~\cite{GBARHbarprod}, $\bar{d}+\text{Ps}\rightarrow \bar{\text{D}} + e^{-}$, a bunch of $10^{4}$ $\bar{\text{d}}$ allows for a production of $0.1~\bar{\text{D}}$ per mixing cycle \cite{GBARantiD}, which would be sufficient to successfully perform spectroscopy measurements.

ASACUSA collaboration intends to carry out laser spectroscopy of antideuteronic helium atoms which should be possible even with very low-intensity antideuteron beams of relatively high energy of a few hundred MeV to test QED and to probe for physics beyond the standard model in a similar way as the antiprotonic helium atom case.


Another possibility is using the antideuterons captured onto atoms and to study the bound-state atomic structure and cascade dynamics of such systems which would give access to unique phenomena never seen before. These could be studied by AEgIS collaboration, or by a dedicated project, EXEQT (EXtrEme Quantum electrodynamics Tests) which is slated to start to pursue these physics cases, first theoretically, and then experimentally when possible. It will use antideuteronic atoms to explore the quantum vacuum in unprecedented field strengths. The antideteuronic atoms will be the first atomic system with a spin 1 bound particle and will be evaluated in a fully relativistic framework. This poses new challenges in QED, like demonstrating renormalizability. On the experimental side, a large solid angle Ge detector system and solid target will be used to see the full cascade after capture by an atom and to observe the breakdown of the antideuteron when the transition energy is larger than the antideuteron binding energy.
The interest of understanding antideuteronic atoms goes way beyond bound state QED. The detection of antideuteronic atoms created by antideuterons in cosmic rays, would be a way to detect these antinuclei, and their presence could be due to dark matter. The GAPS (General AntiParticle Spectrometer) or the ASCENT (A SuperConducting ENergetic x-ray Telescope) using high-altitude balloons will look for antideuteronic and anti-He exotic atoms. Those projects would certainly benefit from better knowledge of antideuteronic atoms spectra and cascade.

\subsection{Antiproton Annihilation and Collision Experiments Using Continuous Beam}
%

If the AD or ELENA machine is upgraded with its own slow extraction beamline, the ASACUSA collaboration plans to study the Pontecorvo reaction via antiproton annihilation on~$^3\mathrm{He}$. The case of three nucleons was not studied in previous measurements at LEAR. At our targeted $10^{-8}$ sensitivity, this would allow one to distinguish between the ``fireball'' and ``rescattering'' models of the nuclear reaction $\bar{p}+\,^3\mathrm{He} \rightarrow n + p$. To further advance our understanding of antiproton-nucleus interactions, we propose to measure annihilation cross-sections at 100 keV using the ELENA decelerator in mini-bunch mode. Depending on future capabilities of the AD, we may extend these studies to 5.3 MeV and above, targeting both annihilation and elastic scattering cross-sections to provide new insights into nuclear structure and fundamental interactions. We will also use our ultra-slow continuous beam for antimatter wave interferometry with material gratings and emulsion detectors, as was recently done in the QUPLAS experiment with positrons. This is both a basic physics experiment and a preparatory step towards the study of the Aharonov-Bohm effect with antiprotons.

\subsection{Low energy antineutrons}

Antinucleon-nucleon interactions were explored at CERN-LEAR, with antineutron beams of momenta above $50\,\mathrm{MeV}/c$. Lower energy antineutrons down to $9\,\mathrm{MeV}/c$  can be produced through a charge-exchange reaction ($\bar{p}p\to\bar{n}n$) with a $300\,\mathrm{MeV}/c$ antiproton beam from the AD \cite{Filippi2025}. It is considered as the possibility to study unresolved problems of low-energy antineutron dynamics, and unveil the origins of a $p\bar{p}$ enhancement and $X$ resonances near the $p\bar{p}$ threshold, as observed in the BESIII experiment.


\newpage
\begin{multicols}{2}
\bibliographystyle{science}
{\footnotesize\bibliography{main}}
\end{multicols}

\appendix
\section*{Appendix}
The Appendix contains the full author list of contributors from each collaboration, listed in alphabetical order.

\section{AEgIS Contributors}

\author{
\begin{center}        
    M.~Auzins\textsuperscript{1}, 
    B.~Bergmann\textsuperscript{2}, 
    P.~Burian\textsuperscript{2},
    R.~S.~Brusa\textsuperscript{3,4}, 
    A.~Camper\textsuperscript{5},
    R.~Caravita\textsuperscript{4},
    F.~Castelli\textsuperscript{6,7},
    G.~Cerchiari\textsuperscript{8,9},
    R.~Ciury\l{}o\textsuperscript{10}, 
    G.~Consolati\textsuperscript{6,11},
    M.~Doser\textsuperscript{12}, 
    M.~Germann\textsuperscript{12}, 
    L.~T.~Gloggler\textsuperscript{12}, 
    L.~Graczykowski\textsuperscript{13}, 
    M.~Grosbart\textsuperscript{12}, 
    F.~Guatieri\textsuperscript{14,3,4}, 
    F.~P.~Gustafsson\textsuperscript{12}, 
    S.~Haider\textsuperscript{12},
    C.~Hugenschmidt\textsuperscript{14}, 
    M.~A.~Janik\textsuperscript{13}, 
    G.~Kasprowicz\textsuperscript{13}, 
    K.~Kempny\textsuperscript{13}, 
    G.~Khatri\textsuperscript{12},
    L.~Klosowski\textsuperscript{10}, 
    G.~Kornakov\textsuperscript{13}, 
    S.~Mariazzi\textsuperscript{3,4},
    P.~Moskal\textsuperscript{15}, 
    D.~Pecak\textsuperscript{16}, 
    L.~Penasa\textsuperscript{3,4}, 
    V.~Petracek\textsuperscript{2}, 
    M.~Piwinski\textsuperscript{10},
    S.~Pospisil\textsuperscript{2},
    F.~Prelz\textsuperscript{6}, 
    B.~S.~Rawat\textsuperscript{17,18},
    B.~Rienacker\textsuperscript{17}, 
    V.~Rodin\textsuperscript{17}, 
    H.~Sandaker\textsuperscript{5},
    S.~Sharma\textsuperscript{15},
    P.~Smolyanskiy\textsuperscript{2}, 
    T.~Sowinski\textsuperscript{16},
    M.~Volponi\textsuperscript{12}, 
    C.~P.~Welsch\textsuperscript{17,18}, 
    M.~Zawada\textsuperscript{10}, 
    N.~Zurlo\textsuperscript{19,20}
        \\ \vspace{0.5cm}
  on behalf of the AEgIS Collaboration\\ \vspace{0.5cm}
\end{center}
    \textsuperscript{1} University of Latvia, Latvia \\
    \textsuperscript{2} Czech Technical University in Prague, Czech Republic \\
    \textsuperscript{3} University of Trento, Italy \\
    \textsuperscript{4} TIFPA/INFN Trento, Italy \\
    \textsuperscript{5} University of Oslo, Norway \\
    \textsuperscript{6} INFN Milano, Italy \\
    \textsuperscript{7} University of Milano, Italy \\
    \textsuperscript{8} University of Siegen, Germany \\
    \textsuperscript{9} Universität Innsbruck, Austria \\
    \textsuperscript{10} Nicolaus Copernicus University in Toru\'n, Poland \\
    \textsuperscript{11} Politecnico di Milano, Italy \\
    \textsuperscript{12} CERN, Switzerland \\
    \textsuperscript{13} Warsaw University of Technology, Poland \\
    \textsuperscript{14} Technical University of Munich, Germany \\
    \textsuperscript{15} Jagiellonian University, Poland \\
    \textsuperscript{16} Polish Academy of Sciences, Warsaw, Poland \\
    \textsuperscript{17} University of Liverpool, United Kingdom \\
    \textsuperscript{18} The Cockcroft Institute, United Kingdom \\
    \textsuperscript{19} INFN Pavia, Italy \\
    \textsuperscript{20} University of Brescia, Italy \\
    
}   

\section{ALPHA Contributors}

\author{
\begin{center}        
        A. Cridland Mathad\textsuperscript{1},
        J. S. Hangst\textsuperscript{2}
        \\ \vspace{0.5cm}
  on behalf of the ALPHA Collaboration\\ \vspace{0.5cm}
\end{center}
      \textsuperscript{1} CERN, Switzerland \\
      \textsuperscript{2} Aarhus Universitet, Denmark \\}      

\section{ASACUSA Contributors}

\author{
\begin{center}
    C.~Amsler\textsuperscript{1},
    M.N.~Bayo\textsuperscript{2,3}, 
    H.~Breuker\textsuperscript{4}, 
    M.~Bumbar\textsuperscript{5,6}, 
    M. Cerwenka\textsuperscript{1}, 
    A. Dax\textsuperscript{7},
    R. Ferragut\textsuperscript{2,3},
    A.~Forsyth Daneri\textsuperscript{8},
    M. Giammarchi\textsuperscript{3}, 
    A. Gligorova\textsuperscript{6},
    T.~Higuchi\textsuperscript{9}, 
    M.~Hori\textsuperscript{8,10},
    E.~D.~Hunter\textsuperscript{5,8},
    K.~Imai\textsuperscript{8},
    V.~Kraxberger\textsuperscript{1}, 
    N.~Kuroda\textsuperscript{11},
    A.~Lanz\textsuperscript{12}, 
    M.~Leali\textsuperscript{13,14},
    G.~Maero\textsuperscript{3,15},
    C.~Malbrunot\textsuperscript{16},
    V.~Mascagna\textsuperscript{3,15},
    Y.~Matsuda\textsuperscript{11},
    D.~J.~Murtagh\textsuperscript{1}, 
    M.~Rome\textsuperscript{3,15}, 
    G.~Roncoli\textsuperscript{3,15}, 
    R.~E.~Sheldon\textsuperscript{1}, 
    M.~C.~Simon\textsuperscript{1}, 
    M.~Tajima\textsuperscript{8,17}, 
    V.~Toso\textsuperscript{13,14}, 
    U.~Uggerh{\o}j\textsuperscript{18}, 
    S.~Ulmer\textsuperscript{4,19}, 
    L.~Venturel\-li\textsuperscript{13,14}, 
    E.~Wid\-mann\textsuperscript{1}, 
    Y.~Yamazaki\textsuperscript{4}
        \\ \vspace{0.5cm}
  on behalf of the ASACUSA Collaboration\\ \vspace{0.5cm}
\end{center}
      \textsuperscript{1} Stefan Meyer Institute, Austria\\
      \textsuperscript{2} Politechnico di Milano, Italy \\
      \textsuperscript{3} INFN Milano, Italy \\
      \textsuperscript{4} RIKEN, Japan \\
      \textsuperscript{5} CERN, Switzerland \\
      \textsuperscript{6} University of Vienna, Austria\\
      \textsuperscript{7} Paul Scherrer Institute, Switzerland \\
      \textsuperscript{8} Imperial College London, United Kingdom \\      
      \textsuperscript{9} Kyoto University, Japan \\
      \textsuperscript{10} Max-Planck-Institut f\"{u}r Quantenoptik, Germany\\
      \textsuperscript{11} University of Tokyo, Japan \\
      \textsuperscript{12} University College London, United Kingdom \\
      \textsuperscript{13} Universit\`a degli Studi di Brescia, Italy \\
      \textsuperscript{14} INFN Pavia, Italy \\
      \textsuperscript{15} Università degli Studi di Milano, Italy \\
      \textsuperscript{16} TRIUMF, Canada \\
      \textsuperscript{17} Japan Synchrotron Radiation Research Institute, Japan \\
      \textsuperscript{18} Aarhus University, Denmark \\
      \textsuperscript{19} Heinrich-Heine-Universität Düsseldorf, Germany
    }

\section{BASE Contributors}

\author{
\begin{center}
        K. Blaum \textsuperscript{1},
        B. M. Latacz\textsuperscript{2},
        Y Matsuda\textsuperscript{3}, 
        C. Ospelkaus\textsuperscript{4,5}, 
        W. Quint\textsuperscript{6},
        J. Walz\textsuperscript{7, 8},
        A. Soter\textsuperscript{9},
        C.~Smorra\textsuperscript{10},
        S. Ulmer\textsuperscript{10,11}
\\ \vspace{0.5cm}
  on behalf of the BASE Collaboration\\ \vspace{0.5cm}
\end{center}
      \textsuperscript{1} Max-Planck-Institut f{\"u}r Kernphysik, Germany\\
      \textsuperscript{2} CERN, Switzerland \\
      \textsuperscript{3} University of Tokyo, Japan\\
      \textsuperscript{4} Leibniz Universit{\"a}t Hannover, Germany\\
      \textsuperscript{5} Physikalisch-Technische Bundesanstalt, Germany \\
      \textsuperscript{6} GSI-Helmholtzzentrum f{\"u}r Schwerionenforschung, Germany \\
      \textsuperscript{7} Johannes Gutenberg-Universit{\"a}t Mainz, Germany \\
      \textsuperscript{8} Helmholtz-Institut Mainz, Germany \\
      \textsuperscript{9} ETH-Z{\"u}rich, Switzerland\\
      \textsuperscript{10} Heinrich-Heine-Universität Düsseldorf, Germany\\
      \textsuperscript{11} RIKEN, Japan}      
      
\newpage
\section{GBAR Contributors}

\author{
\begin{center}
    P. Comini\textsuperscript{1},
    P. Crivelli\textsuperscript{2},
    P. Indelicato\textsuperscript{3},
    N. Kuroda\textsuperscript{4},
    B. Mansoulie\textsuperscript{1},
    P. Perez\textsuperscript{1}
\\ \vspace{0.5cm}
  on behalf of the GBAR Collaboration\\ \vspace{0.5cm}
\end{center}
    \textsuperscript{1} IRFU, CEA, Universite Paris-Saclay, France \\
    \textsuperscript{2} ETH-Z{\"u}rich, Switzerland \\
    \textsuperscript{3} LKB, Sorbonne Universite, France \\
    \textsuperscript{4} University of Tokyo, Japan\\
    }
    
\section{PUMA Contributors}

\author{
\begin{center}
    O. Boine-Frankenheim\textsuperscript{1},
    G. Hupin\textsuperscript{2},
    P. Indelicato\textsuperscript{3},
    M. Kowalska\textsuperscript{4},
    R. Lazauskas\textsuperscript{5},
    A.~Obertelli\textsuperscript{1},
    N.~Paul\textsuperscript{3},
    F. Wienholtz\textsuperscript{1}
\\ \vspace{0.5cm}
  on behalf of the PUMA Collaboration\\ \vspace{0.5cm}
\end{center}
    \textsuperscript{1} Technische Universität Darmstadt, Germany \\
    \textsuperscript{2} Université Paris-Saclay, CNRS/IN2P3, France \\
    \textsuperscript{3} LKB, Sorbonne Université, France \\
    \textsuperscript{4} CERN, Switzerland \\
    \textsuperscript{5} Institut pluridisciplinaire Hubert Curien, CNRS/IN2P3, Université de Strasbourg, France}

\section{External Contributors}

\author{
\begin{center}
        L. Fabbietti\textsuperscript{1},
        H. Fujioka\textsuperscript{2},    
        P. Gasik \textsuperscript{3},
        S. Schiller\textsuperscript{4}
        \\
 \vspace{0.5cm}
\end{center}
      \textsuperscript{1} Technische Universität München, Germany\\
      \textsuperscript{2} Institute of Science Tokyo, Japan\\
      \textsuperscript{3} GSI Helmholtzzentrum für Schwerionenforschung, Germany\\
      \textsuperscript{4} Heinrich-Heine-Universität Düsseldorf, Germany}

\end{document}